\documentclass[twocolumn,amsmath,amssymb]{revtex4}

\usepackage{graphicx,epsf,here}
\usepackage{dcolumn}
\usepackage{bm}

\begin{document}

\title{Abrikosov flux-lines in two-band superconductors with mixed dimensionality}
\author{K. Tanaka$^{1}$ and M. Eschrig$^{2}$} 
\address{
$^{1}$Department of Physics and Engineering Physics, University of Saskatchewan,
116 Science Place, Saskatoon, Saskatchewan, Canada S7N 5E2 \\
$^{2}$Institut f\"ur Theoretische Festk\"orperphysik and DFG-Center for Functional Nanostructures,
Universit\"at Karlsruhe, D-76128 Karlsruhe, Germany}
\begin{abstract}
We study vortex structure in a two-band superconductor, in which one band is ballistic and 
quasi-two-dimensional (2D), and the other is diffusive and three-dimensional (3D).
A circular cell approximation of the vortex lattice within the quasiclassical theory of
superconductivity is applied to a recently developed model appropriate for such a two-band system
[Tanaka {\it et al} 2006 {\it Phys. Rev.} B {\bf 73} 220501(R);
Tanaka {\it et al} 2007 {\it Phys. Rev.} B {\bf 75} 214512].
We assume that superconductivity in the 3D diffusive band is ``weak'', i.e., mostly induced, 
as is the case in MgB$_2$.  Hybridization with the ``weak'' 3D diffusive band has significant and 
intriguing influence on the electronic structure of the ``strong'' 2D ballistic band.
In particular, the Coulomb repulsion and the diffusivity in the ``weak'' band 
enhance suppression of the order parameter and enlargement of the vortex core
by magnetic field in the ``strong'' band, resulting in reduced critical temperature and field.
Moreover, increased diffusivity in the ``weak'' band can result in
an upward curvature of the upper critical field near the transition temperature.
A particularly interesting feature found in our model is the appearance of additional bound states 
at the gap edge in the ``strong'' ballistic band, which are absent in the single-band case.
Furthermore, coupling with the ``weak'' diffusive band leads to reduced
band gaps and van Hove singularities of energy bands of the vortex lattice 
in the ``strong'' ballistic band.
We find these intriguing features for parameter values appropriate for MgB$_2$.
\end{abstract}

\maketitle

\section{Introduction}

Vortices are complex objects even in conventional type-II superconductors, and 
vortex structure can reveal underlying physics of the pairing interaction.
The 40 K superconductor MgB$_2$ \cite{nagamatsu01} 
is the best material discovered so far for studying multiple-band superconductivity.
It can be well described by a two-band model that consists of
the ``strong'' quasi-2D $\sigma$ band (energy gap $\approx 7.2$ meV) 
and the ``weak'' 3D $\pi$ band (energy gap $\approx 2.3$ meV), 
and there is evidence of induced superconductivity 
\cite{schmidt02,eskildsen02,geerk05,giubileo06,klein1}
and strong impurity scattering in the $\pi$ band.
By tunneling along the $c$ axis, Eskildsen {\it et al} have 
probed the vortex core structure in the $\pi$ band, and have 
found that the local density of states (LDOS) is completely flat
as a function of energy at the vortex centre, 
with no signature of localized states \cite{eskildsen02}.
Moreover, the core size as measured by a decay length of the zero-bias LDOS was found to be
much larger than expected from $H_{c2}$.
The existence of two length scales in the vortex lattice has also been suggested by the
$\mu$SR measurement \cite{serventi04}. The thermal conductivity measurement \cite{sologubenko1}
has detected a rapid increase of delocalized quasiparticles for field well below $H_{c2}$.
Unusually large vortex core size and the extended nature of quasiparticle motion 
have also been found in the two-band superconductor NbSe$_2$
\cite{sonier,sonier99,miller00,boaknin03,callaghan05,salman07}.
There is the possibility of multiband superconductivity in a number of other materials
\cite{guritanu04,measson04,seyfarth05,mukhopadhyay05,kasahara07,nakajima08,hunte08},
and understanding the electronic structure in the vortex state of multiband superconductors
is of great interest.

Theoretically, the vortex state in a two-band superconductor
has been studied assuming both bands to be in the clean \cite{nakai02}
and dirty \cite{koshelev03} limit.
With superconductivity (mostly) induced in one band, it has been found that
there are two different length scales associated with the two bands,
and that the averaged zero-bias LDOS increases rapidly as a function of field
in the ``weak'' band. 
Neither of these models, however, are suitable for describing many MgB$_2$ samples,
in which the $\sigma$ and $\pi$ bands are in the ballistic and diffusive limit, respectively.
We have recently formulated a unique two-band model, 
in which one band (``$\sigma$ band'') is ballistic and the other (``$\pi$ band'') is diffusive,
and there is little interband impurity scattering \cite{tanaka06,tanaka07}.
This picture is appropriate for describing MgB$_2$, in which 
two-band superconductivity is retained even in ``dirty'' samples. 
(See extensive references in ref.~\cite{tanaka07}.)
Having MgB$_2$ in mind, in our model it is assumed that
the $\sigma$ and $\pi$ bands are quasi-2D and 3D, respectively, and superconductivity is
mostly induced in the $\pi$ band through the pairing interaction with the $\sigma$ band.
We have examined the electronic structure around an isolated vortex and have found that
the zero-bias LDOS in the $\pi$ band can have a decay length much larger than in the $\sigma$ band.
A particularly intriguing feature that emerges in our model is the possible existence of bound states
at the gap edge in the ballistic band, in addition to the well-known Caroli-de Gennes Matricon
(CdM) bound states \cite{cdm}, in the vortex core.

In this work, we extend our two-band model to describe the vortex lattice and investigate 
the effects of induced superconductivity and impurities in the ``weak'' diffusive band
on the electronic properties in the mixed state.
We have found the presence of highly delocalized quasiparticles for relatively weak field,
in agreement with the experiment on MgB$_2$ \cite{eskildsen02,sologubenko1}
and the earlier theoretical works \cite{nakai02,koshelev03}.
Furthermore, band gaps and van Hove singularities of energy bands of the vortex lattice
in the ``strong'' ballistic band can be reduced by coupling with the ``weak'' diffusive band.
There exist gap-edge bound states in the vortex core in the ballistic band -- that solely arise 
from coupling to the diffusive band -- also in the vortex lattice.

\section{Formulation}
We utilize a recently developed model appropriate for a system
with coupled ballistic and diffusive bands \cite{tanaka06,tanaka07}.
Both the ballistic and diffusive limits can be described within the quasiclassical theory 
of superconductivity, in which all the physical information is contained
in the quasiclassical Green function, or propagator.
We write the propagator in band $\alpha$ as 
$\hat{g}_\alpha(\epsilon, \vec{p}_{F\alpha}, \vec{R})$, where
$\epsilon$ is the quasiparticle energy measured from the chemical potential, 
$\vec{p}_{F\alpha}$ the quasiparticle momentum on the Fermi surface corresponding to band $\alpha $,
and $\vec{R}$ the spatial coordinate.
The hat refers to the 2$\times$2 matrix structure in the Nambu-Gor'kov particle-hole space.  
In the clean $\sigma$ band, $\hat g_\sigma(\epsilon, \vec{p}_{F\sigma}, \vec{R})$ 
satisfies the Eilenberger equation \cite{larkin68,eilenberger},
\begin{equation}
\left[ \epsilon \hat \tau_3 - \hat \Delta_\sigma 
,\; \hat g_\sigma \right]
+i \hbar \vec{v}_{F\sigma} \cdot {\nabla }  \hat g_\sigma
= \hat 0,
\label{eil}
\end{equation}
where $\vec{v}_{F\sigma}$ is the Fermi velocity and $\hat \Delta_\sigma$
the (spatially varying) order parameter. The three Pauli matrices
are denoted by $\hat \tau_i$, $i=1,2,3$, and $[...,... ]$ denotes the commutator.
Motivated by the Fermi surface of MgB$_2$, we assume a cylindrical Fermi surface and treat 
the $\sigma$ band as quasi-two-dimensional.  The $\sigma$-band coherence length is defined as
$\xi_{\sigma}=\hbar v_{F\sigma}/2\pi T_c$, where $T_c$ is the transition temperature, 
and used as length unit.

We assume that the $\pi$ band is in the diffusive limit. 
In the presence of strong impurity scattering, the momentum dependence of the
quasiclassical Green function is averaged out, and 
the equation of motion for the resulting propagator
$\hat g_\pi(\epsilon,\vec{R})$ reduces to the
Usadel equation \cite{usadel70},
\begin{equation}
\label{usadeleq}
\left[ \epsilon \hat \tau_3 - \hat \Delta_\pi ,\; \hat g_\pi \right] 
+ 
{\nabla }
\frac{\hbar\mathbb{D}}{\pi}
(\hat g_\pi {\nabla } \hat g_\pi )
= \hat 0,
\end{equation}
with the diffusion constant tensor $\mathbb{D}$.
Throughout this work, we assume an isotropic tensor $\mathbb{D}_{ij}=D\delta_{ij}$
and define the $\pi$-band coherence length $\xi_\pi=\sqrt{\hbar D/2\pi T_c}$.
Both ballistic and diffusive propagators are normalized according to
$\hat g_\sigma^2=\hat g_\pi^2=-\pi^2 \hat 1$ \cite{larkin68}.

The quasiparticles in different bands are assumed to be coupled only through the pairing interaction.
Selfconsistency for the spatially varying order parameters in the two bands is achieved through
the coupled gap equations,
\begin{equation}
\Delta_\alpha (\vec{R}) = \sum_{\beta} V_{\alpha\beta} N_{F\beta}
{\cal F}_\beta (\vec{R}),
\label{gapeq}
\end{equation}
where $\alpha,\beta \in \{\sigma,\pi\}$, and
$\hat\Delta_\alpha=\hat\tau_1~\rm{Re}~\Delta_\alpha-\hat\tau_2~\rm{Im}~\Delta_\alpha$.
The coupling matrix $V_{\alpha\beta}$ determines
the pairing interaction, $N_{F\beta}$ is the Fermi-surface density of
states on band $\beta$, and
\begin{eqnarray}
{\cal F}_{\sigma }(\vec{R})&\equiv&
\int_{-\epsilon_c}^{\epsilon_c} {d\epsilon\over 2 \pi i}\,
\langle f_\sigma (\epsilon , \vec{p}_{F\sigma},\vec{R}) \rangle_{\vec{p}_{F\sigma}}
\, {\rm tanh}\Biggl({\epsilon\over 2 T}\Biggr),
\label{F1} \nonumber \\
{\cal F}_{\pi }(\vec{R})&\equiv&
\int_{-\epsilon_c}^{\epsilon_c} {d\epsilon\over 2 \pi i}\,
f_\pi (\epsilon, \vec{R} )
\, {\rm tanh}\Biggl({\epsilon\over 2 T}\Biggr).
\label{F2}
\end{eqnarray}
Here $f_\alpha$ is the upper off-diagonal (1,2) element of the matrix
propagator $\hat g_\alpha$, and $\epsilon_c$ is a cutoff energy (that
will be ultimately eliminated as discussed below).
The Fermi surface average over the $\sigma$-band is denoted by
$\langle\cdots\rangle_{\vec{p}_{F\sigma}}$.
In this work, we consider isotropic $s$-wave coupling as is the case for MgB$_2$.

To study the mixed state, we introduce a circular cell approximation, meaning
that we simulate the vortex unit cell by a circle.
Other suggestions for circular cell approximations have been 
introduced in refs.~\cite{kramer74} and \cite{pesch74} for the Usadel and 
Eilenberger equation, respectively.
We assume strong type-II superconductivity, in which case
the spatial variation of magnetic field within the unit cell
can be neglected for fields not too close to $H_{c1}$. 
This is the case for MgB$_2$.
We assume that magnetic field is in the $-z$ direction
(taking into account that the electron charge $e<0$, this gives a positive phase winding).
The vector potential is then given by
\begin{equation}
\vec{A}(\vec{r})= -\frac{H_0}{2} r \vec{e}_{\phi },
\end{equation}
with $\vec{H}_0= -H_0 \vec{e}_z=\nabla \times \vec{A}$ and $r=\sqrt{x^2+y^2}$.
The radius of the vortex unit cell is determined by the fact that one
flux quantum penetrates the unit cell, 
\begin{equation}
H_0\pi r_c^2= \Phi_0,
\label{rctoh}
\end{equation}
with $\Phi_0=hc/2|e|=\pi \hbar c/ |e|$. Thus, 
\begin{equation}
\frac{2e}{\hbar c}A_\phi (r)= \frac{r}{r_c^2}.
\end{equation}
This can be used directly in both the Eilenberger and the Usadel equation.
In both cases, the vector potential can be incorporated by the replacement
\begin{equation}
\nabla_i \hat X \to  \hat \partial_i \hat X \equiv \nabla_i \hat X 
-i \left[ \frac{e}{\hbar c} \hat \tau_3 A_i , \hat X \right]_\circ,
\end{equation}
where the $\circ $ symbol involves a time convolution if $X$ is time-dependent;
otherwise it simply implies matrix multiplication.
Note that in the diffusive case this also affects the expression
for the current density.
This can be obtained by the requirement of local gauge invariance that
the local gauge transformation
\begin{eqnarray}
\hat g &\to & e^{i \chi \hat \tau_3} \circ \hat g \circ e^{-i \chi \hat \tau_3}
\\
\vec{A} &\to & \vec{A} + \frac{\hbar c }{e} \nabla \chi \\
\Phi &\to & \Phi -\frac{\hbar }{e} \partial_t \chi
\end{eqnarray}
with any $\chi (\vec{r},t)$
should leave the basic equations of motion (transport equations) invariant
(in our case there is no time dependence, hence no scalar potential $\Phi$).
The vector and scalar potentials can be gauged {\it locally} away, by
writing down the equations at any point $\vec{R}$ with 
$\nabla \chi = -\frac{e}{\hbar c}\vec{A}$ and $\partial_t \chi = 
\frac{e}{\hbar} \Phi $. 
Thus, the equations with potentials can be obtained by
observing that
\begin{eqnarray}
&&\nabla
\big(e^{i \chi \hat \tau_3} \circ \hat g \circ e^{-i \chi \hat \tau_3}\big)=
\nonumber \\
&&e^{i \chi \hat \tau_3} \circ 
\big(
\nabla \hat g 
+i \left[ \nabla \chi \hat \tau_3 , \hat g \right]_\circ
\big)
\circ e^{-i \chi \hat \tau_3}.
\end{eqnarray}

Now, we use the Riccati parametrization of the Green functions for 
both the Eilenberger \cite{nagato93,schopohl,matthias0,matthias1} and 
Usadel \cite{matthias2} equations:
\begin{equation}
\hat g_\alpha = -\; \frac{i \pi }{1+\gamma_\alpha \tilde \gamma_\alpha }
\left(
\begin{array}{cc}
1-\gamma_\alpha \tilde \gamma_\alpha  & 2\gamma_\alpha \\
2\tilde \gamma_\alpha & \gamma_\alpha \tilde \gamma_\alpha -1
\end{array}\right),
\end{equation}
where $\tilde \gamma_\sigma (\epsilon, \vec{p}_{F\sigma}, \vec{R})=
\gamma^\ast_\sigma (-\epsilon^\ast, -\vec{p}_{F\sigma}, \vec{R})$ and
$\tilde \gamma_\pi (\epsilon, \vec{R})=
\gamma^\ast_\pi (-\epsilon^\ast, \vec{R})$.

Because the transport equations for the Riccati amplitudes must also be
invariant under any local gauge transformation, we can write them down
immediately by noting that the above gauge transformation means
\begin{eqnarray}
\gamma &\to & e^{i \chi } \circ \gamma  \circ e^{i \chi }\\
\tilde \gamma &\to & e^{-i \chi } \circ \tilde \gamma  \circ e^{-i \chi },
\end{eqnarray}
or if we have no time dependence,
\begin{eqnarray}
\gamma &\to & e^{2i \chi } \gamma  \\
\tilde \gamma &\to & e^{-2i \chi } \tilde \gamma .
\end{eqnarray}
Similarly, $\Delta \to e^{2i \chi } \Delta $ and 
$\Delta^\ast \to e^{-2i \chi } \Delta^\ast $.
We thus obtain for the clean $\sigma$ band,
\begin{eqnarray}
\Delta_\sigma +
2 \epsilon \, \gamma_\sigma +
\Delta_\sigma^\ast \gamma_\sigma^2 +
i \hbar v_{f\sigma,i} \partial_i \gamma_{\sigma } = 0,
\label{eileng}
\end{eqnarray}
with $\partial_i= (\nabla_i + 2i\nabla \chi ) = (\nabla_i -i\frac{2e}{\hbar c}A_i)$,
and
\begin{eqnarray}
\Delta_\sigma^\ast -
2 \epsilon \, \tilde \gamma_\sigma +
\Delta_\sigma \tilde \gamma_\sigma^2 +
i \hbar v_{f\sigma,i} \tilde \partial_i \tilde \gamma_{\sigma } = 0,
\label{eilengt}
\end{eqnarray}
with $\tilde \partial_i= (\nabla_i - 2i\nabla \chi ) = (\nabla_i +i\frac{2e}{\hbar c}A_i)$.
For the dirty $\pi$ band, we have \cite{matthias2}
\begin{equation}
\Delta_\pi +
2 \epsilon \, \gamma_\pi +
\Delta_\pi^\ast \gamma_\pi^2  -
i \hbar D \left( \partial_i \partial_i \gamma_\pi - \frac{2\tilde \gamma_\pi (\partial_i \gamma_\pi )^2
}{1+\gamma_\pi \tilde \gamma_\pi } \right) = 0,
\label{usa1}
\end{equation}
and
\begin{equation}
\Delta_\pi^\ast -
2 \epsilon \, \tilde \gamma_\pi +
\Delta_\pi \tilde \gamma_\pi^2  +
i \hbar D \left( \tilde \partial_i \tilde \partial_i \tilde \gamma_\pi - 
\frac{2 \gamma_\pi (\tilde \partial_i \tilde \gamma_\pi )^2
}{1+\gamma_\pi \tilde \gamma_\pi } \right) = 0.
\label{usa2}
\end{equation}
We consider a vortex extending in $z$ direction, with its centre situated at $x=y=0$ for each $z$.
We write $(x,y)= r (\cos \phi , \sin \phi )$ and $(v_{Fx},v_{Fy})=
v_F (\cos \theta , \sin \theta )$, and 
$\Delta = | \Delta (r)| e^{i\phi }$, $\gamma = \gamma (\theta; \phi, r) e^{i\phi }$, and
$\tilde \gamma = \tilde \gamma (\theta; \phi, r) e^{-i\phi }$.
Then we can use the above equations with the quantities $| \Delta (r)|$, 
$\gamma (\theta; \phi, r)$, and $\tilde \gamma (\theta; \phi, r)$, and with the replacements 
$ \partial \to 
(\nabla -i\frac{2e}{\hbar c}A_\phi (r) \vec{e}_\phi +i\vec{e}_\phi/r )$ and 
$\tilde \partial \to (\nabla +i\frac{2e}{\hbar c}A_\phi(r)\vec{e}_\phi 
-i\vec{e}_\phi/r )$.
Here we have used $\nabla \phi = \vec{e}_\phi/r $. 

To obtain the Riccati amplitudes in the 2D ballistic band, we must first solve for
the boundary values along the unit cell circle, from which to integrate the Riccati equations
along a given trajectory.
To simulate the phase change across a unit cell boundary,
we impose the boundary condition (here we omit the subscript $\sigma$),
\begin{equation}
\gamma_{in}(\theta ; \phi, r_c)=-\gamma_{out}(\theta ; \phi+\pi, r_c)\,,
\label{eilenbc}
\end{equation}
where $in$ and $out$ refer to incoming and outgoing trajectories, respectively,
and similarly for $\tilde \gamma$.
For each trajectory we solve for the boundary value $\gamma_{in}$ selfconsistently.

For the 3D diffusive band, we use the symmetry
\begin{eqnarray}
\gamma_\pi ( \phi ,r ) &=& \gamma_\pi (0,r) e^{i\phi }\\
\tilde \gamma_\pi ( \phi ,r ) &=& \tilde \gamma_\pi (0,r) e^{-i\phi }.
\end{eqnarray}
Then equations~(\ref{usa1}) and (\ref{usa2}) reduce to 
the following dimensionless equations
for $\gamma_\pi(0,r)=\gamma_\pi(r)\equiv \gamma_\pi$
and $\tilde \gamma_\pi(0,r)=\tilde \gamma_\pi(r)\equiv \tilde \gamma_\pi$:
\begin{widetext}
\begin{equation}
|\Delta_\pi |(1+\gamma_\pi^2) + 2 \epsilon \, \gamma_\pi -
i \left(\frac{\xi_\pi}{\xi_\sigma }\right)^2 \left( 
\partial^2_r \gamma_\pi + \frac{1}{r} \partial_r \gamma_\pi
-\left( \frac{r}{r_c^2} - \frac{1}{r} \right)^2 \gamma_\pi
\frac{1-\gamma_\pi \tilde \gamma_\pi}{1+\gamma_\pi \tilde \gamma_\pi }
-\frac{2\tilde \gamma_\pi (\partial_r \gamma_\pi )^2 
}{1+\gamma_\pi \tilde \gamma_\pi } \right) = 0,
\label{usadelg}
\end{equation}
\begin{equation}
|\Delta_\pi | (1+\tilde \gamma_\pi^2)
-2 \epsilon \, \tilde \gamma_\pi 
+i \left(\frac{\xi_\pi}{\xi_\sigma }\right)^2 \left(
\partial^2_r \tilde \gamma_\pi + \frac{1}{r} \partial_r \tilde \gamma_\pi
-\left( \frac{r}{r_c^2} - \frac{1}{r} \right)^2 \tilde \gamma_\pi
\frac{1-\gamma_\pi \tilde \gamma_\pi}{1+\gamma_\pi \tilde \gamma_\pi }
-\frac{2 \gamma_\pi (\partial_r \tilde \gamma_\pi )^2
}{1+\gamma_\pi \tilde \gamma_\pi } \right) = 0,
\label{usadelgt}
\end{equation}
\end{widetext}
The boundary condition in the dirty case is that
the first derivative of $\gamma_\pi(r)$ and $\tilde \gamma_\pi(r)$ is zero at the cell boundary. 

We solve for the order parameters by diagonalizing the (cutoff dependent) interactions
and thus transforming the gap equations (\ref{gapeq}) into the form
\begin{equation}
\left(\begin{array}{c}\Delta^{(0)}\\ \Delta^{(1)}\end{array}\right)
= \left(\begin{array}{cc} \lambda^{(0)} & 0\\ 0 & \lambda^{(1)}\end{array}\right)
\left(\begin{array}{c}{\cal F}^{(0)}\\ {\cal F}^{(1)}\end{array}\right)\,,
\label{gapeqd}
\end{equation}
where ${\cal F}^{(0)}$ and ${\cal F}^{(1)}$
are the anomalous amplitudes in the diagonal basis.
The larger eigenvalue, say, $\lambda^{(0)}$,
determines $T_c$ and can be eliminated together with $\epsilon_c$.
The smaller eigenvalue $\lambda^{(1)}$ can be parametrized by
the cutoff-independent quantity
\begin{equation}
\Lambda={\lambda^{(0)}\lambda^{(1)}\over \lambda^{(0)}-\lambda^{(1)}}\,.
\end{equation}
We solve the set of equations (\ref{eileng},\ref{eilengt},\ref{usadelg},\ref{usadelgt})
along with the gap equations for the order parameters selfconsistently.

After selfconsistency has been achieved, the 
LDOS in each band can be calculated by
\begin{eqnarray}
N_{\sigma }(\epsilon,\vec{R})/N_{F\sigma}&=&-
~{\rm Im}~\langle g_\sigma (\epsilon , \vec{p}_{F\sigma},\vec{R})\rangle_{\vec{p}_{F\sigma}} /\pi
,
\nonumber \\
N_{\pi }(\epsilon,\vec{R})/N_{F\pi}&=&-
~{\rm Im}~g_\pi (\epsilon,\vec{R})/\pi,
\label{LDOS}
\end{eqnarray}
where $g_\alpha$ is the upper diagonal (1,1) element of $\hat g_\alpha$.

The current density around the vortex has contributions from
both the $\sigma$ and the $\pi$ band.
The corresponding expressions are 
\begin{eqnarray}
\frac{\vec{j}_{\sigma }(\vec{R})}{2eN_{F\sigma }}&=&
\int_{-\infty}^{\infty} {d\epsilon\over 2\pi }\,
\langle \vec{v}_{F\sigma } {\rm Im}~g_\sigma \rangle_{\vec{p}_{F\sigma}}
\tanh\left( \frac{\epsilon}{2T}\right), \; \nonumber
\\
\frac{\vec{j}_{\pi}(\vec{R})}{2eN_{F\pi}}&=&
\frac{\mathbb{D}}{\pi} \int_{-\infty}^{\infty} {d\epsilon\over 2\pi }\,
{\rm Im}~[f_\pi^{\ast }  {\partial } f_\pi]
\tanh\left( \frac{\epsilon}{2T}\right). \;
\label{CD}
\end{eqnarray}

In our model, the bulk behaviour of the system is completely specified by four material parameters,
$\rho_0$, $n_\pi/n_\sigma$, $T_c$, and $\Lambda$. Here $\rho_0$ is the zero-temperature bulk gap ratio,
$\rho_0=\Delta^{bulk}_\pi/\Delta^{bulk}_\sigma$, and $n_\alpha=N_{F\alpha}/(N_{F\sigma}+N_{F\pi})$.
For MgB$_2$, $\rho_0\approx 0.3$ and $n_\pi/n_\sigma\approx 1-1.2$ 
\cite{eskildsen02,junod01,sologubenko1}.
The zero-temperature gap equations for a homogeneous system relate
$\Lambda$ with $\rho_0$, $n_\pi/n_\sigma$, and the bulk gap ratio near $T_c$ \cite{tanaka07}.
From this relation we find that for MgB$_2$, $\Lambda$ can be positive or negative 
\cite{szabo01,iavarone02,gonnelli02,holanova04} and can be close to zero \cite{iavarone02}.
Negative $\Lambda$ implies that the effective Coulomb interaction dominates over the effective
pairing interaction in the subdominant $\lambda^{(1)}$ channel, and in this case 
superconductivity is purely induced in the $\pi$ band
(with $\Delta_\pi>0$, while $\Delta^{(1)}<0$).

In the presence of inhomogeneity, there is another material parameter, namely 
$\xi_\pi/\xi_\sigma$,
where $\xi_\sigma\simeq 6.8$ nm ($T_c=39$K) for MgB$_2$ \cite{tanaka07}.
As discussed in the next section, to reproduce the experimental data of ref.~\cite{eskildsen02}
we find $\xi_\pi/\xi_\sigma\approx 2$, for $\rho_0=0.3$ and $n_\pi/n_\sigma=1$.
The condition for the $\pi$ band to be in the dirty limit so that the Usadel
equation is applicable is $\xi_\pi/\xi_\sigma < 5$ for MgB$_2$ \cite{tanaka07}.
We present results for $\rho_0=0.1$, 0.3, 0.5;  $\xi_\pi/\xi_\sigma=1$, 3, 5;
$n_\pi/n_\sigma=1$, 1.2; and  $\Lambda=-0.1$, 0, 0.1; and discuss the effects of induced
superconductivity and hybridization of the diffusive and ballistic bands.
For a given cell radius $r_c$, the corresponding field strength is given 
by equation~(\ref{rctoh}) in units of $\hbar c/|e|\xi_\sigma^2$, which is about 13 T for MgB$_2$.

\section{Results}
\subsection{Order parameter}

\begin{figure}
\begin{minipage}{\columnwidth}
\includegraphics[width=0.8\columnwidth]{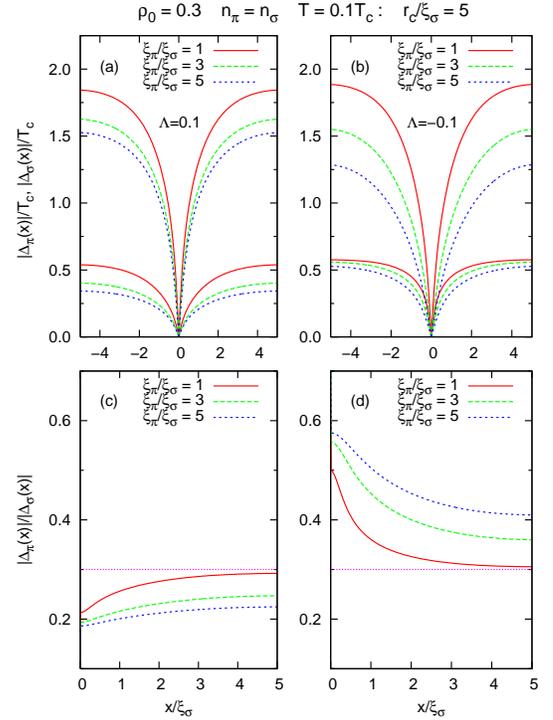}
\end{minipage}
\caption{Order parameters $|\Delta_\pi(x)|$ and $|\Delta_\sigma(x)|$ and their ratio
as a function of coordinate $x$ along a path through the vortex centre
for $\xi_\pi/\xi_\sigma=1,3,5$, $\rho_0=0.3$, $n_\pi=n_\sigma$, $T=0.1T_c$, $r_c/\xi_\sigma=5$,
for $\Lambda=0.1$ (a,c) and -0.1 (b,d).
}
\label{or1}
\end{figure}

\begin{figure}
\begin{minipage}{\columnwidth}
\includegraphics[width=0.8\columnwidth]{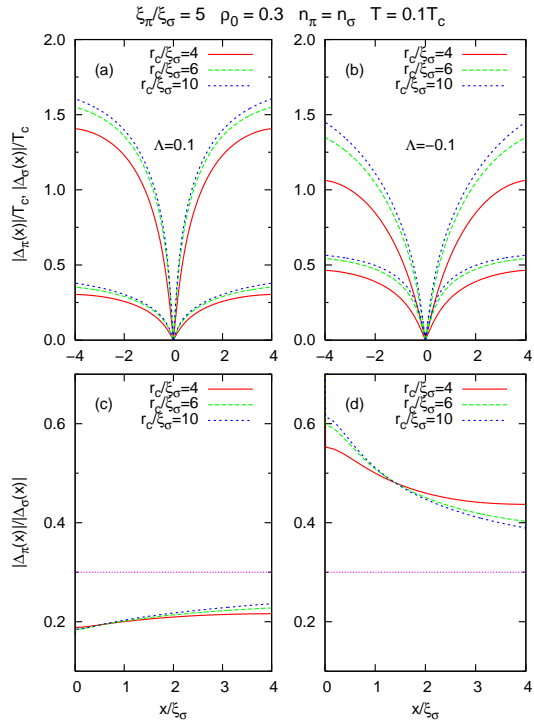}
\end{minipage}
\caption{Order parameters $|\Delta_\pi(x)|$ and $|\Delta_\sigma(x)|$ and their ratio
as a function of coordinate $x$ along a path through the vortex centre
for $\xi_\pi/\xi_\sigma=5$, $\rho_0=0.3$, $n_\pi=n_\sigma$, $T=0.1T_c$, $r_c/\xi_\sigma=4,6,10$, 
for $\Lambda=0.1$ (a,c) and -0.1 (b,d).
}
\label{or2}
\end{figure}

In figure~\ref{or1} we show the order parameter magnitudes in the two bands (a,b) and their ratio
(c,d) as a function of coordinate $x$ along a path through the vortex centre
for $\rho_0=0.3$, $n_\pi=n_\sigma$, and 
$T=0.1T_c$, for various values of the coherence length ratio
$\xi_\pi/\xi_\sigma$; for $\Lambda=0.1$ (a,c) and -0.1 (b,d).
The unit cell radius $r_c/\xi_\sigma=5$ corresponds roughly to 0.5 T for MgB$_2$.
The vortex structure is affected significantly by the Coulomb interaction and 
the impurity scattering rate in the $\pi$ band.  
For $\Lambda > 0$, $|\Delta_\pi(x)/\Delta_\sigma(x)|$ 
is smaller than the bulk gap ratio $\rho_0$ in the vortex core.
While for $\xi_\pi/\xi_\sigma=1$ the ratio recovers more or less to the bulk value at the
cell boundary, as $\xi_\pi/\xi_\sigma$ increases,
$|\Delta_\pi(x)|$ is suppressed more strongly
and $|\Delta_\pi(x)/\Delta_\sigma(x)|$ is smaller than $\rho_0$ in the entire unit cell 
(for $\xi_\pi\sim r_c$).
Furthermore, through coupling with the dirty $\pi$ band, suppression of 
the $\sigma$-band order parameter by magnetic field is also enhanced for larger $\xi_\pi/\xi_\sigma$.
This suppression of the order parameter 
and enlargement of the core area in the $\sigma$ band
with increasing $\xi_\pi/\xi_\sigma$ are more drastic when
the Coulomb repulsion is dominant in the $\pi$ band ($\Lambda < 0$).
Interestingly, in this case, the depletion of $|\Delta_\pi(x)|$ is relatively small
and $|\Delta_\pi(x)/\Delta_\sigma(x)|$ is substantially larger than $\rho_0$,
especially for larger $\xi_\pi/\xi_\sigma$.

\begin{figure}
\begin{minipage}{\columnwidth}
\includegraphics[width=0.8\columnwidth]{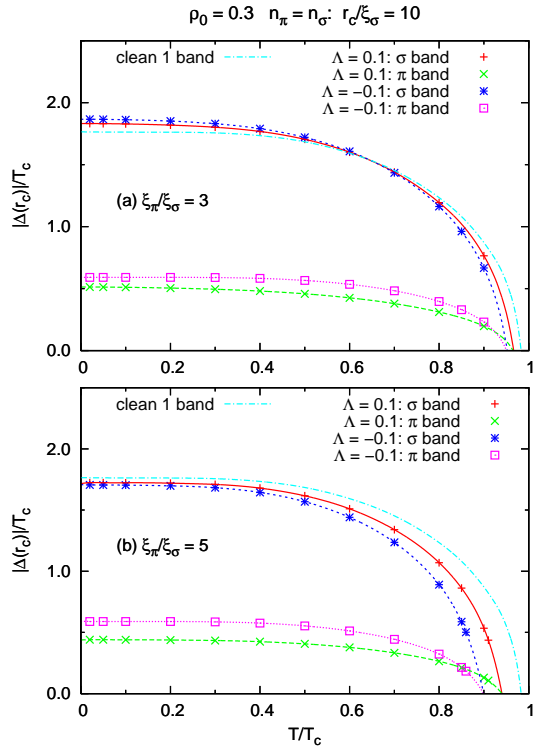}
\end{minipage}
\caption{Order parameter magnitude at the vortex unit cell boundary in the two bands
as a function of temperature $T$ for $\rho_0=0.3$, $n_\pi=n_\sigma$, $\Lambda=\pm 0.1$, $r_c/\xi_\sigma=10$, 
for (a) $\xi_\pi/\xi_\sigma=3$ and (b) 5.  The result for a single clean band is also shown.
}
\label{dt1}
\end{figure}

\begin{figure}
\begin{minipage}{\columnwidth}
\includegraphics[width=0.8\columnwidth]{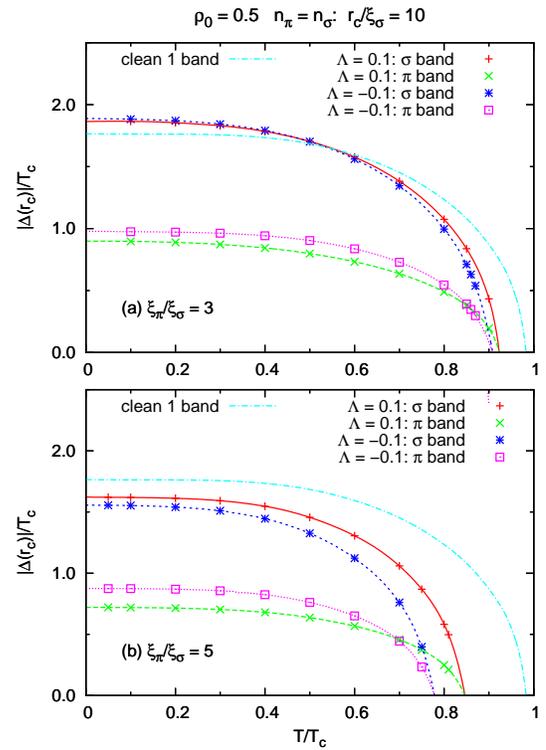}
\end{minipage}
\caption{Same as figure~\ref{dt1} except for $\rho_0=0.5$.
}
\label{dt2}
\end{figure}

Changes in the order parameter magnitudes in the two bands and their ratio as the field strength changes
are illustrated in figure~\ref{or2}, for $\xi_\pi/\xi_\sigma=5$, $\rho_0=0.3$, $n_\pi=n_\sigma$, and 
$T=0.1T_c$, for various values of $r_c$; for $\Lambda=0.1$ (a,c) and -0.1 (b,d).
The result for $r_c/\xi_\sigma=10$ is similar to that for an isolated vortex.
It can be seen clearly that for $\Lambda > 0$, the $\pi$-band order parameter is suppressed
strongly by magnetic field and the core area is enlarged.
In contrast, for $\Lambda < 0$, the depairing effect is more manifest 
in the $\sigma$ band, with $|\Delta_\pi(x)/\Delta_\sigma(x)|$ reaching about two times $\rho_0$
in the vortex centre.  It is clear that, while superconductivity is (mostly) induced in
the $\pi$ band, Coulomb interactions can renormalise substantially
the length scales and the core sizes in the two bands in different ways.

We illustrate the influence of the diffusivity in the $\pi$ band on the critical temperature 
in figures~\ref{dt1} ($\rho_0=0.3$) and \ref{dt2} ($\rho_0=0.5$)
for $n_\pi=n_\sigma$ and $r_c/\xi_\sigma=10$, for (a) $\xi_\pi/\xi_\sigma=3$ and (b) 5.
In these figures the order parameter magnitudes at the cell boundary, $|\Delta(r_c)|$, 
in the two bands are plotted as a function of $T$ for $\Lambda=0.1$ and -0.1, 
along with that for a single clean band.
Points are results obtained by selfconsistent calculation and curves are guides to the eye.
In a single ballistic band, the critical temperature where $|\Delta(r_c)|$ vanishes is slightly
smaller than the zero-field $T_c$.
When the ballistic $\sigma$ band is coupled with the diffusive $\pi$ band, 
the critical temperature is reduced further,
and superconductivity is more suppressed by magnetic field for $\Lambda=-0.1$
than for $\Lambda=0.1$.
This difference is enhanced for larger $\xi_\pi/\xi_\sigma$, as a result of stronger
suppression of the $\sigma$-band order parameter as discussed above.
These effects of the diffusivity and the Coulomb interaction in the $\pi$ band 
can be drastic when the coupling between the two bands is strong,
as demonstrated for $\rho_0=0.5$ in figure~\ref{dt2}.

A quantity to characterise the vortex core structure is the vortex core size defined by
\cite{kramerpesch,kato01}
\begin{equation}
\xi_{c}^{-1} = {\partial \Delta(r=0)\over \partial r}
{1\over \Delta(r=\infty)}\,,
\label{coresize}
\end{equation}
where $r$ is the radial coordinate measured from the vortex centre,
and $\Delta(r=\infty)\equiv \Delta(r_c)$ is the `bulk' order parameter in the vortex lattice.
Around an isolated vortex in a single clean band, the order parameter exhibits 
the KP effect \cite{kramerpesch}, i.e., shrinkage of the vortex core size
as $T$ is lowered, approaching zero in the zero-temperature limit \cite{kato01,hayashi05,gumann06}.
This is due to depopulation of higher-energy bound states in the vortex core.
In an $s$-wave superconductor with nonmagnetic impurities, however, the core size as defined above
saturates as $T$ approaches zero \cite{hayashi05,gumann06}.
This stems from broadening of the bound core states that removes the singular behaviour 
in the spatial variation of the order parameter in the vortex core.
The vortex core shrinking ceases when $k_B T$ becomes smaller than the energy width 
of the zero-energy bound states in the core.

\begin{figure}
\begin{minipage}{\columnwidth}
\includegraphics[width=0.8\columnwidth]{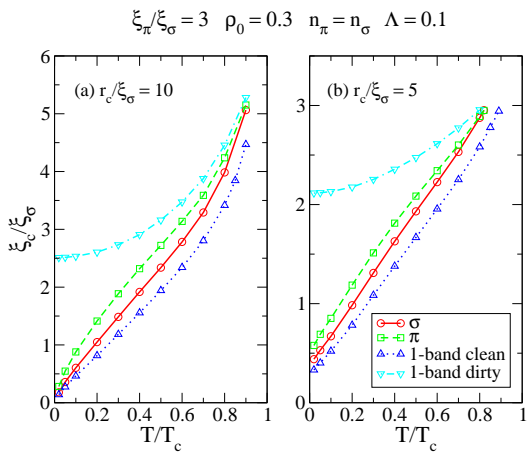}
\end{minipage}
\caption{Vortex core size as defined by equation~(\ref{coresize}) in the two bands as a function of 
temperature $T$ for $\xi_\pi/\xi_\sigma=3$, $\rho_0=0.3$, $n_\pi=n_\sigma$, $\Lambda=0.1$,
for (a) $r_c/\xi_\sigma=10$ and (b) $r_c/\xi_\sigma=5$.  The results for a single clean and dirty band
are also shown.
}
\label{corep}
\end{figure}

\begin{figure}
\begin{minipage}{\columnwidth}
\includegraphics[width=0.8\columnwidth]{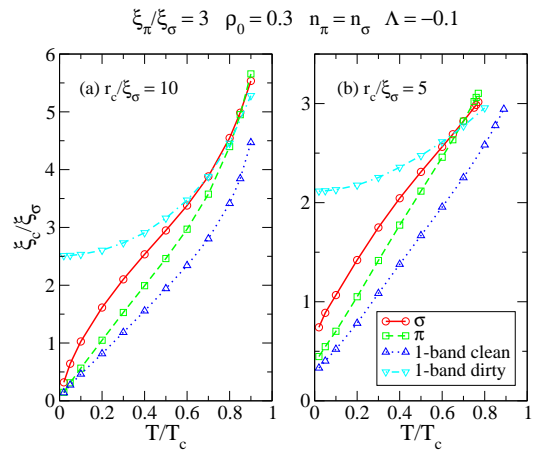}
\end{minipage}
\caption{Same as figure~\ref{corep} except for $\Lambda=-0.1$.
}
\label{corem}
\end{figure}

\begin{figure}
\begin{minipage}{\columnwidth}
\includegraphics[width=0.8\columnwidth]{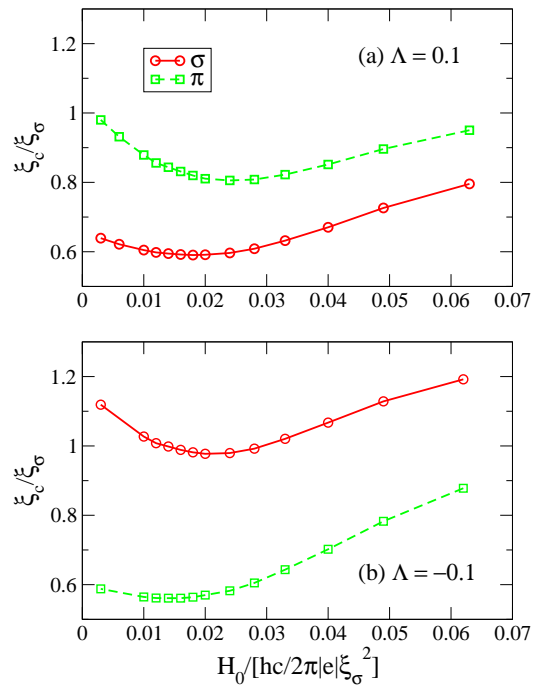}
\end{minipage}
\caption{Vortex core size as defined by equation~(\ref{coresize}) in the two bands as a function of 
field $H_0$ for $\xi_\pi/\xi_\sigma=3$, $\rho_0=0.3$, $n_\pi=n_\sigma$, $T/T_c=0.1$, for
(a) $\Lambda=0.1$ and $\Lambda=-0.1$.
}
\label{coreh}
\end{figure}

When a ballistic band and a diffusive band are coupled, the KP effect is induced in the diffusive
band, as found in refs.~\cite{gumann06,tanaka07} for an isolated vortex.
In the vortex lattice, when vortices are well separated, one finds the KP effect 
in a single clean band as well as in coupled clean and dirty bands.
This is demonstrated in figures~\ref{corep} ($\Lambda=0.1$) and \ref{corem} ($\Lambda=-0.1$),
in which $\xi_c$ in the two bands 
is plotted as a function of $T$ for $\xi_\pi/\xi_\sigma=3$, $\rho_0=0.3$,
and $n_\pi=n_\sigma$, for (a) $r_c/\xi_\sigma=10$ and (b) $r_c/\xi_\sigma=5$.
The core size in the case of a single clean and dirty band is also shown.
As can be seen in figure~\ref{corep}(a), for $\Lambda>0$, 
the core size in the $\pi$ band is larger than that in the $\sigma$ band for all temperature.
The $\pi$-band order parameter exhibits the KP effect also when the Coulomb repulsion dominates
(figure~\ref{corem}(a)), and in this case, $\xi_c$ as defined above is always smaller
in the $\pi$ band than in the $\sigma$ band (except for $T$ very close to the critical temperature).
Also note that, with dominating Coulomb interactions, the $T$-linear behaviour of the KP effect
is better developed in the $\pi$ band than in the $\sigma$ band.

As magnetic field increases and vortices come closer together, the bound state wavefunctions
of neighbouring vortices begin to overlap and form energy bands, and quasiparticles 
can travel through the periodic array of vortices \cite{canel65,klein89,poettinger93,sonier}
(see also references in ref.~\cite{sonier}).
We have performed our calculation for the cell radius $r_c\ge 4$.
For this range of $r_c$, we find that in a single diffusive band, 
the core size $\xi_c$ becomes smaller as $r_c$ decreases for any given temperature.
Such shrinkage of the vortex core as field increases can be understood 
as due to intervortex transfer of quasiparticles \cite{sonier}, 
with higher-energy core bound states turning into extended states.
This is also a trend for a single clean band as well as coupled ballistic and diffusive bands
for relatively high $T$ and relatively large $r_c$.
However, as can be seen in figures~\ref{corep}(b) and \ref{corem}(b), for relatively strong field,
$\xi_c$ is finite in the zero-temperature limit;
since the derivative of the order parameter at the vortex centre remains finite as 
temperature approaches zero.
Thus for a given (low) temperature, $\xi_c$ can increase as a function of field strength
above a certain critical value.
This is illustrated in figure~\ref{coreh}, where $\xi_c$ is plotted as a function of field strength $H_0$
for $T/T_c=0.1$, $\xi_\pi/\xi_\sigma=3$, $\rho_0=0.3$, and $n_\pi=n_\sigma$, for (a) $\Lambda=0.1$
and (b) $\Lambda=-0.1$.
In both bands, as field increases, the core size is reduced initially, but at some critical
strength it starts increasing.
In a single ballistic band, $\xi_c$ behaves similarly as a function of $H_0$:
this is consistent with the work by Miranovi\'c {\it et al} \cite{miranovic04}, who examined
the effects of impurities on the vortex core size by including impurity scattering in
the Eilenberger equation (single band).  They have also found that for very small mean free path,
the core size decreases monotonically as a function of field strength, 
as we find for a single diffusive band.
In the case of two bands, $\xi_c$ in the $\pi$ band has  nonmonotonic behaviour 
as in the $\sigma$ band, and the low-temperature value of $\xi_c$ is larger (smaller) in the
$\pi$ band than in the $\sigma$ band for $\Lambda >0$ ($<0$).

\subsection{Spectral properties of diffusive band}

\begin{figure}
\begin{minipage}{\columnwidth}
\includegraphics[width=0.8\columnwidth]{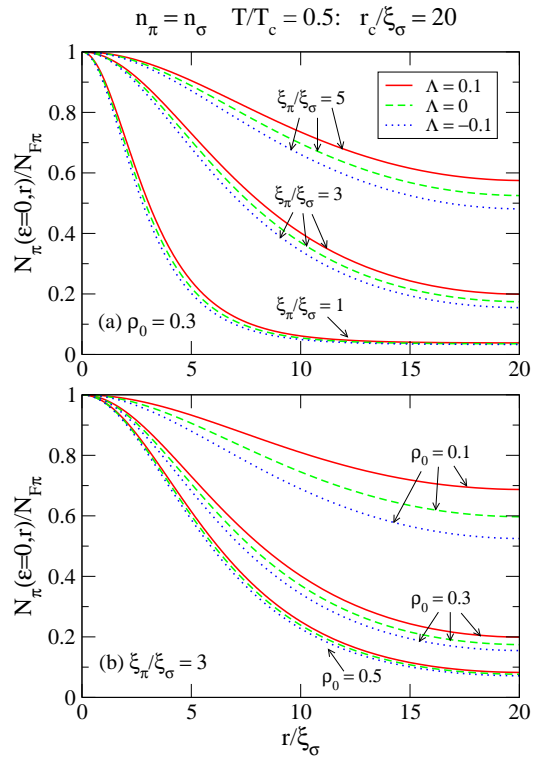}
\end{minipage}
\vspace{0.3cm}
\caption{Zero-bias LDOS in the $\pi$ band as a function of distance $r$ from the vortex centre
for $n_\pi=n_\sigma$, $T/T_c = 0.5$, $\Lambda=0.1,0,-0.1$, $r_c/\xi_\sigma=20$,
for (a) $\rho_0=0.3$, $\xi_\pi/\xi_\sigma=1,3,5$ and (b) $\xi_\pi/\xi_\sigma=3$, $\rho_0=0.1,0.3,0.5$.
}
\label{ldose}
\end{figure}

\begin{figure}
\begin{minipage}{\columnwidth}
\includegraphics[width=0.8\columnwidth]{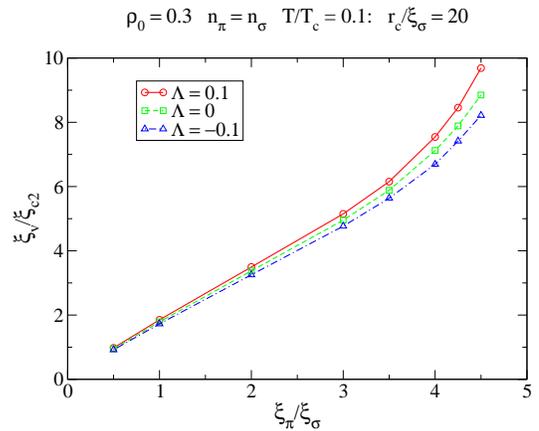}
\end{minipage}
\caption{Half-decay length $\xi_v$ of the zero-bias LDOS in the $\pi$ band as a function of 
$\xi_\pi/\xi_\sigma$
for $\rho_0=0.3$, $n_\pi=n_\sigma$, $T/T_c = 0.1$, $r_c/\xi_\sigma=20$, $\Lambda=0.1,0,-0.1$.
For MgB$_2$, $\xi_{c2}=10$ nm and the observed $\xi_v\approx 30$ nm corresponds to
$\xi_\pi/\xi_\sigma\approx 2$ in our model.
}
\label{xiv}
\end{figure}

In the diffusive $\pi$ band, the LDOS is completely flat as a function of energy at the vortex centre,
as found for an isolated vortex \cite{tanaka06,tanaka07} and for coupled diffusive bands
\cite{koshelev03}.  This is in agreement with the observation with STM \cite{eskildsen02}.
The vortex core size as measured by a decay length of the zero-bias LDOS has also been probed by STM
\cite{eskildsen02}.
When superconductivity is (mostly) induced in a ``weak'' band in a coupled two-band system,
such a decay length can be much larger in the ``weak'' band than in the ``strong'' band
\cite{nakai02,koshelev03,tanaka06,tanaka07}.
We find that in the vortex state, there is little change in
the zero-bias LDOS as a function of $r$ in the $\sigma$ band from that for an isolated vortex,
except for rather small $r_c$.
In contrast, as $r_c$ is reduced, the $\pi$ band starts exhibiting substantial 
overlap of quasiparticle states across the cell boundary for relatively large $r_c$.
Furthermore, the LDOS decay length in the $\pi$ band strongly depends on 
$\rho_0$ and $\xi_\pi/\xi_\sigma$.
In figure~\ref{ldose} we present the zero-bias LDOS in the $\pi$ band as a function of distance $r$
from the vortex centre for $n_\pi=n_\sigma$, $T/T_c = 0.5$, $\Lambda=0.1,0,-0.1$, 
and $r_c/\xi_\sigma=20$;
for (a) $\rho_0=0.3$ for various values of $\xi_\pi/\xi_\sigma$ and (b) $\xi_\pi/\xi_\sigma=3$ for
several values of $\rho_0$.
For parameter values appropriate for MgB$_2$ 
($\xi_\pi/\xi_\sigma\approx 2$, $\rho_0=0.3$; see below), 
the overlap of core states is not negligible already for $r_c/\xi_\sigma=20$.
For a fixed $\rho_0$, the diffusivity in the $\pi$ band (larger $\xi_\pi/\xi_\sigma$) 
enhances intervortex transfer of $\pi$-band quasiparticles.
For a given  $\xi_\pi/\xi_\sigma$, the weaker the induced superconductivity,
the stronger the extension and overlap of quasiparticle states.
The Coulomb repulsion in the $\pi$ band (negative $\Lambda$) can enhance these effects further:
e.g., for the parameter set for figure~\ref{ldose}(b), the dependence of the LDOS
on $\Lambda$ is significant for $\rho_0=0.1$.

To characterise the core size,
we plot in figure~\ref{xiv} the half-decay length of the zero-bias LDOS in the $\pi$ band, $\xi_v$,
in units of $\xi_{c2}\equiv 10$ nm (the Ginzburg-Landau coherence length in MgB$_2$ \cite{eskildsen02})
as a function of $\xi_\pi/\xi_\sigma$
for $\rho_0=0.3$, $n_\pi=n_\sigma$, $T/T_c = 0.1$, and $r_c/\xi_\sigma=20$.
The $\xi_v$ is a linear function of $\xi_\pi/\xi_\sigma$ for $\xi_\pi/\xi_\sigma\lesssim 3$ 
in this example, and increases rapidly as $\xi_\pi/\xi_\sigma$ increases further.
Also for larger $\xi_\pi/\xi_\sigma$, its dependence on $\Lambda$ becomes noticeable.
For MgB$_2$, the experimental value of $\xi_v\approx 30$ nm corresponds to $\xi_\pi/\xi_\sigma\approx 2$
in our model.
In ref.~\cite{koshelev03} such a plot of $\xi_v$ was made for coupled dirty bands for 
$\xi_\pi/\xi_\sigma<2.5$: overall $\xi_v$ in this case is slightly smaller than our $\xi_v$.

\subsection{Spectral properties of ballistic band}

\begin{figure}
\begin{minipage}{\columnwidth}
\includegraphics[width=0.8\columnwidth]{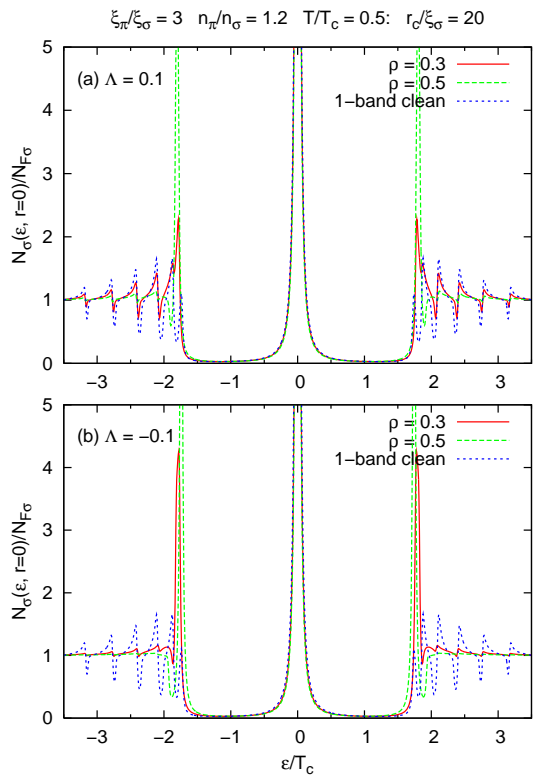}
\end{minipage}
\caption{Vortex centre spectra in the $\sigma$ band as a function of energy $\epsilon$
for $\rho_0=0.3$, 0.5, $\xi_\pi/\xi_\sigma=3$, $n_\pi/n_\sigma=1.2$, $T/T_c = 0.5$, $r_c/\xi_\sigma=20$,
for (a) $\Lambda=0.1$ and (b) -0.1.  The result for a single ballistic band is also shown.
}
\label{dos1}
\end{figure}

\begin{figure}
\begin{minipage}{\columnwidth}
\includegraphics[width=0.8\columnwidth]{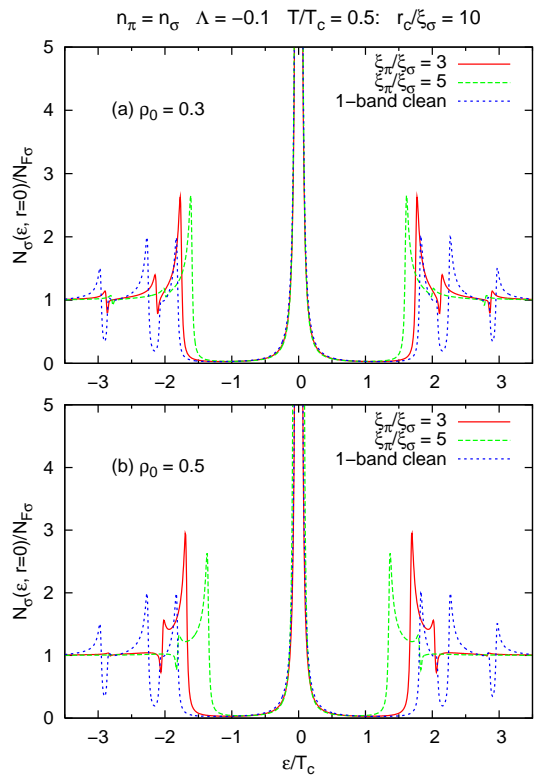}
\end{minipage}
\caption{Vortex centre spectra in the $\sigma$ band as a function of energy $\epsilon$
for $\xi_\pi/\xi_\sigma=3,5$, $n_\pi=n_\sigma$, $\Lambda=-0.1$, $T/T_c = 0.5$, $r_c/\xi_\sigma=10$,
for (a) $\rho_0=0.3$ and (b) 0.5.  The result for a single ballistic band is also shown.
}
\label{dos2}
\end{figure}

A particularly interesting feature found in our model of coupled ballistic and diffusive bands
is that, through coupling with the ``weak'' diffusive band,
there can be additional bound states at the gap edge in the ``strong'' ballistic band.
We find such bound states for an isolated vortex \cite{tanaka06,tanaka07}
as well as for the vortex lattice.
In figure~\ref{dos1} the LDOS at the vortex centre in the $\sigma$ band
is plotted as a function of energy $\epsilon$ for $\rho_0=0.3$, 0.5, 
$\xi_\pi/\xi_\sigma=3$, $n_\pi/n_\sigma=1.2$, $T/T_c = 0.5$, and $r_c/\xi_\sigma=20$;
for (a) $\Lambda=0.1$ and (b) -0.1.
We present in figure~\ref{dos2} the LDOS for $\xi_\pi/\xi_\sigma=3$, 5,
$n_\pi=n_\sigma$, $\Lambda=-0.1$, $T/T_c = 0.5$, and $r_c/\xi_\sigma=10$; 
for (a) $\rho_0=0.3$ and (b) 0.5.
The result for a single ballistic band is also shown in these figures.
The peak at $\epsilon=0$ reflects the CdM bound states that arise from repeated Andreev scattering
from the order parameter in the vortex core.
In the core of an isolated vortex in a single clean band, the spectrum shows neither coherence
peak nor additional bound state at the gap edge.
In the vortex lattice,
as quasiparticles travel in the periodic pairing potential,
the continuum energy levels above the energy gap are grouped into bands
and van Hove singularities appear at the band edges \cite{poettinger93}.
This can be seen clearly for a single ballistic band in figures~\ref{dos1} and \ref{dos2}:
in the latter, the band widths are larger due to reduced lattice spacing.

When the ballistic $\sigma$ band is coupled with the diffusive $\pi$ band,
the vortex core area is enlarged and the periodic pairing potential becomes effectively weaker.
As a result, band gaps and van Hove singularities are reduced as can be seen
in figures~\ref{dos1} and \ref{dos2}.  
At the same time, extra bound states tend to appear at the gap edge.
These features become prominent for larger $\rho_0$ (0.3 vs. 0.5 in figure~\ref{dos1})
and $\xi_\pi/\xi_\sigma$ (3 vs. 5 in figure~\ref{dos2}), and are more enhanced for negative $\Lambda$
than for positive $\Lambda$ (figure~\ref{dos1}).
In most of the examples shown in these figures, the spectrum above the energy gap is more or less
continuous. Furthermore, in the two-band results presented in figure~\ref{dos2}(b),
the energy gap is reduced substantially and a band of gap edge bound states is formed.

\subsection{Current density}

\begin{figure}
\begin{minipage}{\columnwidth}
\includegraphics[width=0.8\columnwidth]{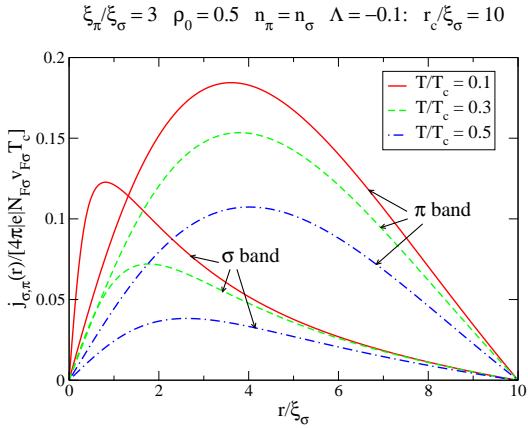}
\end{minipage}
\caption{Current density in the two bands as a function of distance $r$ from the vortex centre 
for $\xi_\pi/\xi_\sigma=3$, $\rho_0=0.5$, $n_\pi=n_\sigma$, $\Lambda=-0.1$, $r_c/\xi_\sigma=10$,
for $T/T_c=0.1,0.3,0.5$.
}
\label{current}
\end{figure}

\begin{figure}
\begin{minipage}{\columnwidth}
\includegraphics[width=0.8\columnwidth]{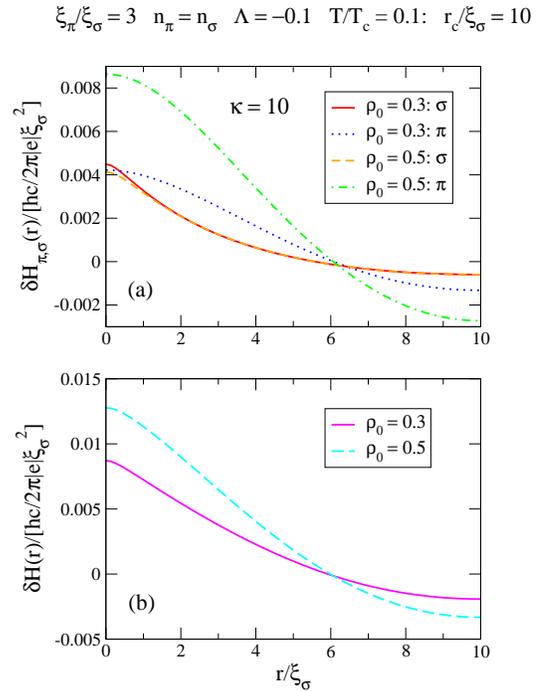}
\end{minipage}
\vspace{0.1cm}
\caption{Magnetic field distribution $\delta H=H-H_0$ as a function of distance $r$ from the
vortex centre for $\xi_\pi/\xi_\sigma=3$, $n_\pi=n_\sigma$, $\Lambda=-0.1$, $T/T_c=0.1$,
$r_c/\xi_\sigma=10$, for $\rho_0=0.3,0.5$; (a) partial contribution from each band and 
(b) total field distribution. 
The $\kappa\equiv \sqrt{8\pi e^2 v_{F\sigma}^2 N_{F\sigma}/c^2}/\xi_\sigma$ that comes into
the Maxwell equation is taken to be 10 ($\kappa\approx 14$ for MgB$_2$).
}
\label{field}
\end{figure}

In figure~\ref{current} the current densities $j(r)$ in the two bands as a function of
distance $r$ from the vortex centre are shown
for $\xi_\pi/\xi_\sigma=3$, $\rho_0=0.5$, $n_\pi=n_\sigma$, $\Lambda=-0.1$, and $r_c/\xi_\sigma=10$,
for various temperatures.
The current density contribution from the $\pi$ band can be substantial,
and even dominating as demonstrated in this figure, for larger $\rho_0$ and $\xi_\pi/\xi_\sigma$.
As temperature is lowered, the $\sigma$-band current density exhibits the KP effect
and is more confined around the vortex centre.
For stronger coupling of the two bands, the KP effect becomes manifest also in the $\pi$ band.

Deviation of the magnetic field distribution from the uniform field, 
$\delta H(r)=H(r)-H_0$, can be obtained from $j(r)$ by integrating
the Maxwell equation, where a parameter 
$\kappa\equiv \sqrt{8\pi e^2 v_{F\sigma}^2 N_{F\sigma}/c^2}/\xi_\sigma$ comes in.
For MgB$_2$, $\kappa\approx 14$.
The $\delta H(r)$ as a function of $r$ is illustrated in figure~\ref{field} for
$\xi_\pi/\xi_\sigma=3$, $n_\pi=n_\sigma$, $\Lambda=-0.1$, $T/T_c=0.1$, $r_c/\xi_\sigma=10$, and $\kappa=10$,
for $\rho_0=0.3$ and 0.5. In figure~\ref{field}(a) partial contribution from each band is
shown, while \ref{field}(b) presents the total field distribution.
It can be seen that when coupling of the two bands is increased,
field fluctuations in the $\pi$ band (hence those in the total field) are enhanced significantly, 
while the $\sigma$ band contribution hardly changes.

\subsection{Phase diagram}

\begin{figure}
\begin{minipage}{\columnwidth}
\includegraphics[width=0.8\columnwidth]{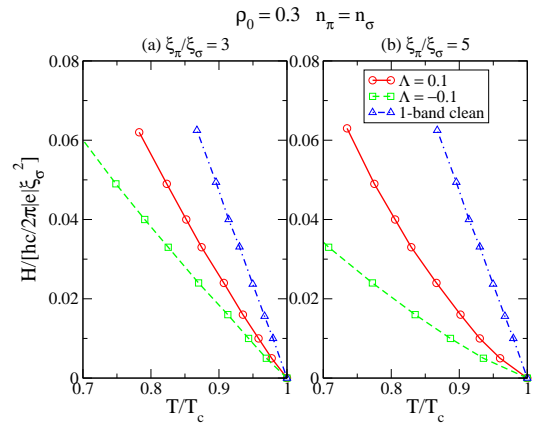}
\end{minipage}
\caption{Upper critical field as a function of temperature $T$ for
$\rho_0=0.3$, $n_\pi=n_\sigma$, $\Lambda=\pm 0.1$, for (a) $\xi_\pi/\xi_\sigma=3$ and (b) 5.
The result for a single clean band is also shown.
The $\hbar c/|e|\xi_\sigma^2\approx 13$ T for MgB$_2$.
}
\label{phase}
\end{figure}

We show in figure~\ref{phase} the upper critical field $H_{c2}(T)$ for $T$ near $T_c$
for $\rho_0=0.3$, $n_\pi=n_\sigma$, and $\Lambda=\pm 0.1$, for (a) $\xi_\pi/\xi_\sigma=3$
and (b) 5, along with the result for a single clean band.
Although $H_{c2}$ is sensitive to the amount of impurities in the sample,
in MgB$_2$ single crystals, $H_{c2}$ for field parallel to the $c$ axis 
has been measured to be $H_{c2}(T=T_c/2)\approx 2$ Tesla in several experiments
\cite{angst02,sologubenko2,zehetmayer02,lyard02,angst03,welp03,shi03,bando04,eisterer05,kim06,klein2}.
This value roughly corresponds to $\Lambda$ somewhere between -0.1 and 0.1 for 
$\xi_\pi/\xi_\sigma=3$, $\rho_0=0.3$, and $n_\pi=n_\sigma$ in our model.
When the $\sigma$ band is coupled with the $\pi$ band, $H_{c2}$ is substantially reduced
from the single-band value, and decreases further when the Coulomb repulsion dominates or
the diffusivity increases in the $\pi$ band.
This is consistent with stronger suppression of the order parameter and enlargement of the
vortex core area in the $\sigma$ band.
Furthermore, for relatively large $\xi_\pi/\xi_\sigma$, 
$H_{c2}(T)$ develops an upward curvature near $T_c$ (figure~\ref{phase}(b)).
This is interesting in light of the theoretical work based on the Eliashberg theory
by Mansor and Carbotte,
who studied the effects of Fermi velocity anisotropy and impurities on $H_{c2}$ \cite{mansor05}.
Their prediction for MgB$_2$ is that, for field in the $c$ direction (where the Fermi velocities
in the two bands are assumed to be the same), $H_{c2}(T)$ exhibits an upward curvature near $T_c$
when the $\pi$ band is clean and the $\sigma$ band is dirty, while it has a quasilinear $T$
dependence for the dirty $\pi$ and the clean $\sigma$ band (as observed in several experiments;
see references in ref.~\cite{mansor05}).
Such an upward curvature in $H_{c2}$ ($\parallel c$) has been observed in some
experiments on MgB$_2$ single crystals \cite{lee01,xu01,kim02}.

\section{Conclusion}

In conclusion, we have studied the effects of induced superconductivity and impurities 
on the electronic structure in the vortex lattice of a two-band superconductor,
in which a ``weak'' 3D diffusive band and a ``strong'' 2D ballistic band 
are hybridized through the pairing interaction.
We have found that the Coulomb repulsion and the diffusivity in the ``weak'' dirty band
enhance suppression of the order parameter and enlargement of the vortex core
by magnetic field in the ``strong'' clean band.
As a result, critical temperature and field 
(where the order parameter at the vortex unit cell boundary vanishes) are reduced,
and increased diffusivity in the dirty band can result in an upward curvature 
of the upper critical field near the transition temperature.
The Kramer-Pesch effect 
arising from thermal depopulation of higher-energy
core bound states tends to disappear as field becomes stronger and 
overlap of quasiparticle states of neighbouring vortices increases.
The zero-bias LDOS in the diffusive band has a decay length much larger than that in
the ballistic band, indicating substantial intervortex transfer of quasiparticles,
and the half-decay length increases significantly as the diffusivity increases.
Coupling with the ``weak'' diffusive band leads to 
smearing of energy bands of the vortex lattice and van Hove singularities at the band edges
in the vortex-core LDOS in the ``strong'' ballistic band.
Furthermore, bound states tend to appear at the gap edge in the ballistic band,
in addition to the well-known Caroli-de Gennes-Matricon bound states.
These effects are enhanced for increased coupling, diffusivity, and Coulomb repulsion.
Finally, the current density contribution in the vortex core and resulting field fluctuations 
in the diffusive band can be substantial, or even dominating, and exhibit the Kramer-Pesch
effect manifest in the current density when coupling with the ballistic band is strong.
We find the above intriguing features in the quasiparticle spectra 
for parameter values appropriate for MgB$_2$.

\section{Acknowledgements}
We thank A. E. Koshelev for helpful discussions and
constructive comments on the manuscript.
The research was supported by the 
Natural Sciences and Engineering Research Council of Canada,
the Canada Foundation for Innovation,
and the Deutsche Forschungsgemeinschaft within the CFN.

\end{document}